\documentclass[12pt]{article}
  \usepackage{amsfonts}
  \usepackage{amsmath}
\usepackage{amssymb}
\usepackage{amscd}
\usepackage[dvips]{graphicx}
\begin{document}

~~
\bigskip
\bigskip
\begin{center}
{\Large {\bf{{{Twist deformation of $l$-conformal Galilei Hopf algebra}}}}}
\end{center}
\bigskip
\bigskip
\bigskip
\begin{center}
{{\large ${\rm {Marcin\;Daszkiewicz}}$}}
\end{center}
\bigskip
\begin{center}
\bigskip

{ ${\rm{Institute\; of\; Theoretical\; Physics}}$}

{ ${\rm{ University\; of\; Wroclaw\; pl.\; Maxa\; Borna\; 9,\;
50-206\; Wroclaw,\; Poland}}$}

{ ${\rm{ e-mail:\; marcin@ift.uni.wroc.pl}}$}

\end{center}
\bigskip
\bigskip
\bigskip
\bigskip
\bigskip
\bigskip
\bigskip
\bigskip
\bigskip
\begin{abstract}
The six Abelian twist-deformations of   $l$-conformal Galilei Hopf algebra are considered. The corresponding twisted
space-times are derived as well.
\end{abstract}
\bigskip
\bigskip
\bigskip
\bigskip
\eject

\section{{{Introduction}}}

The idea to use noncommutative coordinates is quite old - it goes
back to Heisenberg and was firstly formalized by Snyder in
\cite{snyder}. Recently, however,  there were  found new formal
arguments based mainly on Quantum Gravity \cite{2w}, \cite{2a} and
String Theory models \cite{recent}, \cite{string1}, indicating that
space-time at Planck scale  should be noncommutative, i.e. it should
have a quantum nature.

Presently, it is well known, that in accordance with the
Hopf-algebraic classification of all deformations of relativistic
and nonrelativistic symmetries, one can distinguish three
types of quantum spaces \cite{class1}, \cite{class2} (for details see also \cite{nnh}):\\
\\
{ \bf i)} Canonical ($\theta^{\mu\nu}$-deformed) type of quantum space \cite{oeckl}-\cite{dasz1}
\begin{equation}
[\;\hat{ x}_{\mu},\hat{ x}_{\nu}\;] = i\theta_{\mu\nu}\;, \label{noncomm}
\end{equation}
\\
{ \bf ii)} Lie-algebraic modification of classical space-time \cite{dasz1}-\cite{lie1}
\begin{equation}
[\;\hat{ x}_{\mu},\hat{ x}_{\nu}\;] = i\theta_{\mu\nu}^{\rho}\hat{ x}_{\rho}\;,
\label{noncomm1}
\end{equation}
and\\
\\
{ \bf iii)} Quadratic deformation of Minkowski and Galilei  spaces \cite{dasz1}, \cite{lie1}-\cite{paolo}
\begin{equation}
[\;\hat{ x}_{\mu},\hat{ x}_{\nu}\;] = i\theta_{\mu\nu}^{\rho\tau}\hat{
x}_{\rho}\hat{ x}_{\tau}\;, \label{noncomm2}
\end{equation}
with coefficients $\theta_{\mu\nu}$, $\theta_{\mu\nu}^{\rho}$ and  $\theta_{\mu\nu}^{\rho\tau}$ being constants.\\
\\
Besides, it has been demonstrated in \cite{nnh}, that in the case of
so-called N-enlarged Newton-Hooke Hopf algebras
$\,{\mathcal U}^{(N)}_0({ NH}_{\pm})$ the twist deformation
provides the new  space-time noncommutativity of the
form\footnote{$x_0 = ct$.},\footnote{ The discussed space-times have been  defined as the quantum
representation spaces, so-called Hopf modules (see e.g. \cite{oeckl}, \cite{chi}), for quantum N-enlarged
Newton-Hooke Hopf algebras.}
\begin{equation}
{ \bf 4)}\;\;\;\;\;\;\;\;\;[\;t,{ x}_{i}\;] = 0\;\;\;,\;\;\; [\;{ x}_{i},{ x}_{j}\;] = 
if_{\pm}\left(\frac{t}{\tau}\right)\theta_{ij}(x)
\;, \label{nhspace}
\end{equation}
with time-dependent  functions
$$f_+\left(\frac{t}{\tau}\right) =
f\left(\sinh\left(\frac{t}{\tau}\right),\cosh\left(\frac{t}{\tau}\right)\right)\;\;\;,\;\;\;
f_-\left(\frac{t}{\tau}\right) =
f\left(\sin\left(\frac{t}{\tau}\right),\cos\left(\frac{t}{\tau}\right)\right)\;,$$
$\theta_{ij}(x) \sim \theta_{ij} = {\rm const}$ or
$\theta_{ij}(x) \sim \theta_{ij}^{k}x_k$ and  $\tau$ denoting the time scale parameter
 -  the cosmological constant. Moreover,  the  different relations  between all mentioned above quantum spaces ({\bf 1)}, { \bf 2)}, { \bf 3)}
and { \bf 4)}) have been summarized in paper \cite{nnh}.

It should be noted that the  described  above classification can be supplemented by the proper deformations of so-called $l$-conformal Galilei Hopf structure \;${\cal U}^{(l)}_0(\mathcal{G})$ provided in \cite{czar}. In general, a conformal extension of the Galilei Hopf algebra is parametrized by a positive
half integer $l$, which justifies the term $l$-conformal Galilei Hopf structure. Particularly, the instance of
$l=\frac{1}{2}$ (well-known in the literature as the Schroedinger algebra) has been the focus of most studies (for a review see e.g. \cite{przeglad}).
Besides, motivated by current investigation of the nonrelativistic version of the AdS/CFT correspondence, the interest in conformal
Galilei Hopf algebras with $l>\frac{1}{2}$ is growing rapidly in the last time  \cite{pocz}-\cite{kon}.

In this article  we investigate  the Abelian twist deformations of \;${\cal U}^{(l)}_0(\mathcal{G})$ Hopf structure
which provide (as we shall see for a moment)  six types of space-time noncommutativity (see formulas (\ref{rspacetime1})-(\ref{rspacetime6})).
 The motivations for such a kind of  studies are manyfold.  First of all, we construct explicitly the deformation of  quite general (known)
Hopf algebra at nonrelativistic level.
Secondly, we get the completely new quantum (twist-deformed) space-times associated with \;${\cal U}^{(l)}_{\alpha}(\mathcal{G})$ Hopf
structure. Such a result seems to be quite interesting due to the fact, that it extends in natural way the mentioned above classification
of the quantum spaces. Finally, it should be noted  that the obtained results permit to consider the classical as well as quantum particle models defined on new  noncommutative nonrelativistic space-times (\ref{rspacetime1})-(\ref{rspacetime6}).

The paper is organized as follows. In first section we recall  the basic facts concerning $l$-conformal Galilei Hopf algebra \;${\cal U}^{(l)}_0(\mathcal{G})$. The
second section is devoted to its twist deformations and to the derivation of corresponding quantum space-times. The  final remarks are provided in the last section.

\section{{{$l$-conformal Galilei Hopf algebra \;${\cal U}^{(l)}_0(\mathcal{G})$}}}

In this section we recall  basic facts associated with the $l$-conformal Galilei Hopf algebra \;${\cal U}^{(l)}_0(\mathcal{G})$ provided in article \cite{czar}.
Hence, the $l$-conformal Galilei Hopf algebra includes the generators of time translations, dilatations, special conformal transformations, spatial rotations,
spatial translations, Galilei boosts
and accelerations. Denoting the mentioned generators by $H$, $D$, $K$, $M_{ij}$ and $G^{(n)}_i$, respectively, where $i=1,\dots,d$ is a spatial index and $n=0,1,\dots, 2l$,  one can write the following algebraic
\begin{eqnarray}
&&\left[\;M_{ij},M_{kl}\;\right] = -i(\delta_{ik} M_{jl}+\delta_{jl} M_{ik} - \delta_{il} M_{jk}-\delta_{jk} M_{il})\;, \label{algebra1}\\
&~~&  \cr
&&\left[\;M_{ij},G^{(n)}_k\;\right] = -i (\delta_{ik} G^{(n)}_j-\delta_{jk} G^{(n)}_i)\;, \label{algebra2}\\
&~~&  \cr
&&\left[\;K,G^{(n)}_i\;\right] = i (n-2l) G^{(n+1)}_i\;\;\;,\;\;\;
[D,G^{(n)}_i] = i (n-l) G^{(n)}_i\;, \label{algebra3}\\
&~~&  \cr
&&\left[\;H,K\;\right] = 2 i D\;\;\;,\;\;\; \left[\;D,K\;\right] = i K\;,\label{algebra4}\\
&~~&  \cr
&&\left[\;H,D\;\right] = i H\;\;\;,\;\;\; \left[\;H,G^{(n)}_i\;\right]\;=\;i n G^{(n-1)}_i\;,\label{algebra5}
\end{eqnarray}
as well as coalgebraic
\begin{eqnarray}
\Delta_0(a) = a\otimes 1 +1\otimes a\;\;\;,\;\;\;S_{0}(a) =-a\;,\label{cop}
\end{eqnarray}
sectors. It should be also observed that operators $H$, $D$ and $K$ form $so(2,1)$ subalgebra, which is the conformal algebra in one dimension. Besides, one can
notice that the instances of $n=0$ and $n=1$ in $G^{(n)}_i$ correspond to the spatial translations and Galilei boosts respectively, while the
higher values of index $n$ are linked to the accelerations. Finally, it is easy to check that all above generators are represented on the classical space of functions as follows \cite{anton}
\begin{eqnarray}
M_{ij}&\rhd& f(t,\overline{x}) =i\left( x_{i }{\partial_j} -x_{j
}{\partial_i} \right) f(t,\overline{x})\;,\label{a1}\\
H&\rhd& f(t,\overline{x})=i{\partial_t}f(t,\overline{x})\;, \label{a2}\\
G_i^{(n)}&\rhd& f(t,\overline{x})=it^n
{\partial_i}f(t,\overline{x})\;,  \label{a3}\\
D&\rhd& f(t,\overline{x})= i\left(t \partial_t+l x^i \partial_i\right)f(t,\overline{x})\;,\label{a4}\\
K&\rhd& f(t,\overline{x})=i\left(t^2 \partial_t+2 l t x^i \partial_i\right)f(t,\overline{x})\;.\label{a5}
\end{eqnarray}
Of course, for $D=K=0$ and $n=0,1$ we reproduce well-known (ordinary) Galilei Hopf structure $\;{\cal U}_{0}(\mathcal{G})$.

\section{Twist deformations of $l$-conformal Galilei Hopf algebra and the corresponding quantum space-times}

Let us now turn to the twist deformations of the Hopf structure described in pervious section. First of all, in accordance with Drinfeld  twist procedure
\cite{twist1}-\cite{twist3}, the algebraic sector of twisted
$l$-conformal Galilei Hopf algebra $\;{\cal U}^{(l)}_{\alpha}(\mathcal{G})$ remains
undeformed  (see (\ref{algebra1})-(\ref{algebra5})), while
the   coproducts and antipodes  transform as follows (see formula (\ref{cop}))
\begin{equation}
\Delta _{0}(a) \to \Delta _{\alpha }(a) = \mathcal{F}_{\alpha }\circ
\,\Delta _{0}(a)\,\circ \mathcal{F}_{\alpha }^{-1}\;\;\;,\;\;\;
S_{\alpha}(a) =u_{\alpha }\,S_{0}(a)\,u^{-1}_{\alpha }\;,\label{fs}
\end{equation}
with $u_{\alpha }=\sum f_{(1)}S_0(f_{(2)})$ (we use Sweedler's notation
$\mathcal{F}_{\alpha }=\sum f_{(1)}\otimes f_{(2)}$).
Besides, it should be noted, that the twist factor
$\mathcal{F}_{\alpha } \in {\cal U}^{(l)}_{\alpha}(\mathcal{G}) \otimes
{\cal U}^{(l)}_{\alpha}(\mathcal{G})$
satisfies  the classical cocycle condition
\begin{equation}
{\mathcal F}_{{\alpha }12} \cdot(\Delta_{0} \otimes 1) ~{\cal
F}_{\alpha } = {\mathcal F}_{{\alpha }23} \cdot(1\otimes \Delta_{0})
~{\mathcal F}_{{\alpha }}\;, \label{cocyclef}
\end{equation}
and the normalization condition
\begin{equation}
(\epsilon \otimes 1)~{\cal F}_{{\alpha }} = (1 \otimes
\epsilon)~{\cal F}_{{\alpha }} = 1\;, \label{normalizationhh}
\end{equation}
with ${\cal F}_{{\alpha }12} = {\cal F}_{{\alpha }}\otimes 1$ and
${\cal F}_{{\alpha }23} = 1 \otimes {\cal F}_{{\alpha }}$.

It is well known, that the twisted algebra $\;{\cal U}^{(l)}_{\alpha}(\mathcal{G})$ can be described in terms of
so-called classical $r$-matrix $r\in {\cal U}^{(l)}_{\alpha}(\mathcal{G}) \otimes {\cal U}^{(l)}_{\alpha}(\mathcal{G})$,
which satisfies the  classical Yang-Baxter equation (CYBE)
\begin{equation}
[[\;r_{\alpha},r_{\alpha}\;] ] = [\;r_{\alpha 12},r_{\alpha13} +
r_{\alpha 23}\;] + [\;r_{\alpha 13}, r_{\alpha 23}\;] = 0\;,
\label{cybe}
\end{equation}
where   symbol $[[\;\cdot,\cdot\;]]$ denotes the Schouten bracket
and for $r = \sum_{i}a_i\otimes b_i$
$$r_{ 12} = \sum_{i}a_i\otimes b_i\otimes 1\;\;,\;\;r_{ 13} = \sum_{i}a_i\otimes 1\otimes b_i\;\;,\;\;
r_{ 23} = \sum_{i}1\otimes a_i\otimes b_i\;.$$

In this article we consider six types of Abelian twist deformation of $l$-conformal Galilei Hopf algebra, described by the following
$r$-matrices\footnote{$a \wedge b = a \otimes b - b\otimes a.$}
\begin{eqnarray}
r_1 &=&  \frac{1}{2}{\alpha^{ij}_1} G_i^{(n)} \wedge
G_i^{(m)}\;\;\;\;\;\;\;\, [\;\alpha^{ij} = -\alpha^{ji}\;]\;,
\label{rmacierze01}\\
r_2 &=& \alpha_2 G_i^{(n)}
\wedge M_{kl}\;\;\;\;\;\;[\;i,k,l - {\rm fixed},\;\;i \neq
k,l\;]\;,\label{rmacierze02}\\
r_3 &=&  {\alpha_3} H \wedge M_{kl}\;,
\label{rmacierze03}\\
r_4 &=&  {\alpha_4} K \wedge M_{kl}\;,
\label{rmacierze04}\\
r_5 &=&  {\alpha_5} K \wedge
G_i^{(n)}\;\;\;\;\;\;\;\;[\;n=2l\;]\;,
\label{rmacierze05}\\
r_6 &=&  {\alpha_6} D \wedge M_{kl}\;,
\label{rmacierze06}
\end{eqnarray}
where $\alpha^{ij}_1$, $\alpha_2$, $\alpha_3$, ... $\alpha_6$ denote the deformation parameters.
Due to the Abelian character of the above carriers (all of them contain
 the mutually commuting elements of the algebra), the
corresponding twist factors can be obtained in a  standard way
\cite{twist1}-\cite{twist3}, i.e. they take the form
\begin{eqnarray}
{\cal F}_{{a}} = \exp
\left(ir_a\right)\;\;\;;\;\;\;a=1,2,\dots ,6\;.
\label{factors}
\end{eqnarray}

The corresponding quantum space-times are defined as the representation spaces (Hopf modules) for $l$-conformal Galilei Hopf algebra
\;${\cal U}_{\alpha}^{(l)}({{\mathcal{G}}})$, with action of the generators $M_{ij}$, $H$, $G_i^{(n)}$, $K$ and $D$ given by (\ref{a1})-(\ref{a5})
(see e.g. \cite{oeckl}, \cite{chi}). Besides, the $\star$-multiplication of arbitrary two functions covariant under
$\;{\cal U}^{(l)}_{\alpha}(\mathcal{G})$ is defined as follows
\begin{equation}
f(t,\overline{x})\star_a g(t,\overline{x}):=
\omega\circ\left(
 (\mathcal{F}_a)^{-1}\rhd  f(t,\overline{x})\otimes g(t,\overline{x})\right)
 \;,\label{star1}
\end{equation}
where symbol  $\mathcal{F}_a$ denotes the  twist factors (see
(\ref{factors}))   and $\omega\circ\left( a\otimes b\right) =
a\cdot b$. Consequently, we get
\begin{eqnarray}
&{\bf 1.~~~~~~}&[\,t,x_a\,]_{{\star_1}} =0\;\;\;,\;\;\;
[\,x_a,x_b\,]_{{\star}_1}
= i\alpha^{ij}t^{n+m}
 (\delta_{ai}\delta_{bj} -
\delta_{aj}\delta_{bi})\;,\label{rspacetime1}\\
&~~&  \cr
&{\bf 2.~~~~~~}&[\,t,x_a\,]_{{\star}_2} =0\;\label{rspacetime2}\\
&~~&  \cr
&&[\,x_a,x_b\,]_{{\star}_2} =2i\alpha_2 t^n
\left[\;\delta_{ia}(x_k\delta_{bl} - x_{l}\delta_{bk}) -
\delta_{ib}(x_k\delta_{al} -
x_{l}\delta_{ak})\;\right]\;,\nonumber\\
&~~&  \cr
&{\bf 3.~~~~~~}&[\,t,x_a\,]_{{\star}_3} =2i\alpha_3(x_k\delta_{la}-x_l\delta_{ka})\;\;\;,\;\;\;
[\,x_a,x_b\,]_{{\star}_3} =0\;,\label{rspacetime3}\\
&~~&  \cr
&{\bf 4.~~~~~~}&[\,t,x_a\,]_{{\star}_4} =2i\alpha_4t^2(x_k\delta_{la}-x_l\delta_{ka})\;,\label{rspacetime4}\\
&~~&  \cr
&&[\,x_a,x_b\,]_{{\star}_4} =4i\alpha_4 lt
\left[\;x_{a}(x_k\delta_{bl} - x_{l}\delta_{bk}) -
x_{b}(x_k\delta_{al} -
x_{l}\delta_{ak})\;\right]\;,\nonumber\\
&~~&  \cr
&{\bf 5.~~~~~~}&[\,t,x_a\,]_{{\star}_5} =2i\alpha_5t^{2(1+l)}\delta_{ia}\;,\label{rspacetime5}\\
&~~&  \cr
&&[\,x_a,x_b\,]_{{\star}_5} =4i\alpha_5 lt^{2l+1}
(x_{a}\delta_{bi} - x_{b}\delta_{ai}) \;,
\end{eqnarray}
and
\begin{eqnarray}
{\bf 6.~~~~~~~}[\,t,x_a\,]_{{\star}_6} &=&2i\alpha_6t(x_k\delta_{la}-x_l\delta_{ka})\;,\label{rspacetime6}\\
&~~&  \cr
[\,x_a,x_b\,]_{{\star}_6} &=&2i\alpha_6 l
\left[\;x_{a}(x_k\delta_{bl} - x_{l}\delta_{bk}) -
x_{b}(x_k\delta_{al} -
x_{l}\delta_{ak})\;\right]\;,\nonumber
\end{eqnarray}
respectively. It should be noted that three first spaces are the same as in the case of so-called twisted $N$-enlarged Galilei  algebra \cite{nnh}, while the remaining ones
correspond to the conformal sector of considered in present article Hopf structure. Obviously, for   deformation parameters $\alpha^{ij}_1$ and $\alpha_2,\dots , \alpha_6$ approaching
zero the above quantum  space-times  become classical.

\section{Final remarks}

In this article we consider six  Abelian twist-deformations of
$l$-conformal Galilei Hopf algebra $\;{\cal U}^{(l)}_{\alpha}(\mathcal{G})$. The corresponding twisted
space-times are derived as well. It should be noted, however, that  present studies can be extended in various
ways.  First of all, one can find the dual Hopf structures
$\,{\mathcal D}_{\alpha}^{(l)}(G)$ with the use of FRT
procedure \cite{frt} or by canonical quantization of the
corresponding Poisson-Lie structures \cite{poisson}. Besides, as it
was already mentioned in Introduction, one should ask about the
basic dynamical models corresponding to the $l$-conformal space-times
(\ref{rspacetime1})-(\ref{rspacetime6}).   Finally, one can also consider more
complicated (non-Abelian) twist deformations of
$l$-conformal Hopf algebras, i.e. one can find
the twisted coproducts, corresponding noncommutative space-times and
dual Hopf structures. Such  problems are now under consideration.

\section*{Acknowledgments}
The author would like to thank J. Lukierski
for valuable discussions. This paper has been financially  supported  by Polish
NCN grant No 2011/01/B/ST2/03354.

\end{document}